%% file: main.tex
\documentclass[letterpaper, 10 pt, conference]{ieeeconf}  

\IEEEoverridecommandlockouts                              

\usepackage{graphicx}
\usepackage{subcaption}
\usepackage{booktabs}
\usepackage{epsfig} 
\usepackage{amsmath} 
\usepackage{amssymb}  
\usepackage{mathrsfs}
\usepackage{cleveref}
\usepackage{color}
\usepackage{calc}
\usepackage{dsfont}
\usepackage{bm}
\usepackage{float}
\newcommand\figref{Fig.~\ref}

\newtheorem{lemma}{Lemma}
\newtheorem{theorem}{Theorem}

\newtheorem{corollary}{Corollary}
\newtheorem{remark}{Remark}

\usepackage{algorithm} 
\usepackage{algpseudocode} 
\usepackage{xspace}
\usepackage{multirow}

\usepackage{amsfonts} 
\title{\LARGE \bf Towards Data-Driven Model-Free Safety-Critical Control}

\author{Zhe Shen, Yitaek Kim, and Christoffer Sloth 
\thanks{Authors are with the Maersk Mc-Kinney Moller Institute, University of Southern Denmark, Denmark {\tt\small \{zhes,yik,chsl\}@mmmi.sdu.dk}}
}

\begin{document}
\maketitle
\thispagestyle{empty}
\pagestyle{empty}


\begin{abstract}
This paper presents a framework for enabling safe velocity control of general robotic systems using data-driven model-free Control Barrier Functions (CBFs). Model-free CBFs rely on an exponentially stable velocity controller and a design parameter (e.g. $\alpha$ in CBFs); this design parameter depends on the exponential decay rate of the controller. However, in practice, the decay rate is often unavailable, making it non-trivial to use model-free CBFs, as it requires manual tuning for $\alpha$. To address this, a Neural Network is used to learn the Lyapunov function from data, and the maximum decay rate of the system’s built-in velocity controller is subsequently estimated. Furthermore, to integrate the estimated decay rate with model-free CBFs, we derive a probabilistic safety condition that incorporates a confidence bound on the violation rate of the exponential stability condition, using Chernoff bound. This enhances robustness against uncertainties in stability violations. The proposed framework has been tested on a UR5e robot in multiple experimental settings, and its effectiveness in ensuring safe velocity control with model-free CBFs has been demonstrated.
\end{abstract}

\input{introduction}

\input{related_works}
\input{preliminary_knowledges}
\input{neural_network}

\input{proposed_framework}

\input{method}
\input{experiment}

\input{conclusions}

\bibliographystyle{IEEEtran}
\bibliography{bibliography}
\end{document}

%% file: introduction.tex
\section{Introduction}\label{sec:introduction}
Control Barrier Functions (CBFs) have been extensively explored to provide safety guarantees in various applications, including commercial robotics \cite{Singletary2022}. While significant progress has been made in safety-critical controllers, the effectiveness of these controllers in guaranteeing safety is often constrained by the complexity and uncertainty of the system model. This challenge arises because most CBF-based controllers heavily depend on accurate knowledge of the system model. To mitigate this challenge, a novel model-independent method, known as Model-Free CBFs (MF-CBFs) \cite{molnar2021model}, has recently been devised to ensure safety without requiring explicit knowledge of the system dynamics in safe controller design. The fundamental idea behind model-free safety guarantees is based on an exponentially stable velocity tracking controller, which is widely implemented in robotic systems \cite{Spong2005}. 

However, the design of a safe controller using MF-CBFs \cite{molnar2021model} assumes that prior knowledge of the built-in velocity controller is available. Nonetheless, the presence of unknowns in the system model makes implementation challenging, as it heavily relies on heuristic tuning to achieve the desired performance. For instance, the exponential decay rate of the Lyapunov function candidate plays a crucial role in shaping $\alpha$ in a safe velocity controller using model-free CBFs. Despite this obvious practical limitation, to the best of our knowledge, there has not yet been explicitly investigated about how a safe velocity controller from \cite{molnar2021model} can be systematically designed on a built-in velocity tracking controller.
\begin{figure}[t]
    \centering
    \includegraphics[width=1\linewidth]{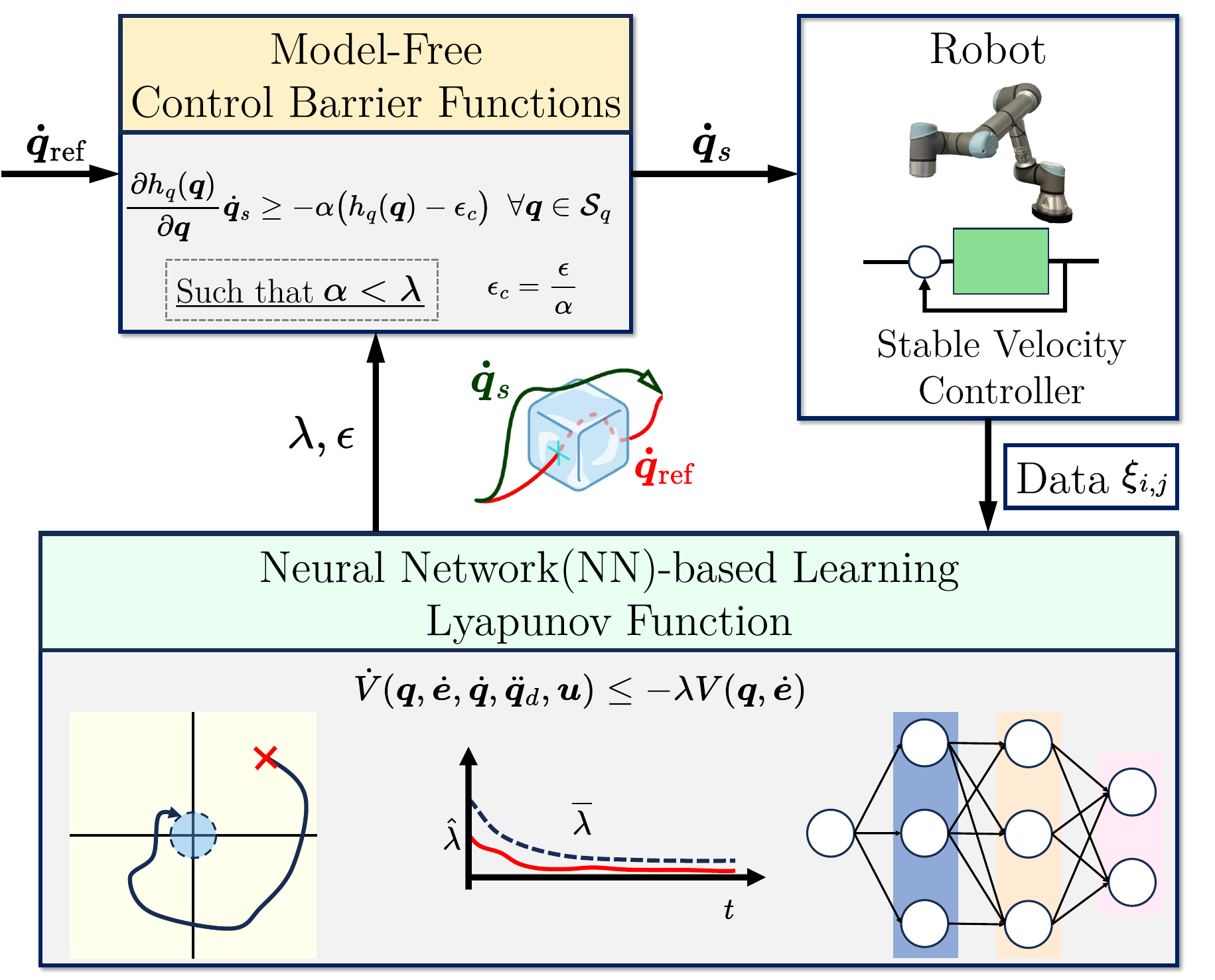}
    \caption{\small{Illustration of the proposed framework for ensuring safety guarantees using Model-Free Control Barrier Functions (MF-CBFs). From data (e.g. joint trajectories), we learn Lyapunov function of a built-in velocity controller in the robot and estimate its maximum decay rate, $\lambda$ and constraint violation parameter, $\epsilon$. Lastly, $\lambda$ and $\epsilon$ are then used to design $\alpha$ in MF-CBFs. The proposed framework accomplishes probabilistic safety guarantees, avoiding manual tuning $\alpha$.}}
    \label{pro:framework}
\end{figure}
\subsection{Contributions}
We propose a data-driven framework for ensuring safety via model-free CBFs, which efficiently accomplishes safe velocity control without requiring accurate model knowledge or heuristic tuning to achieve the desired control performance. 

To this end, we use a Neural Network (NN) to learn a Lyapunov function associated with the built-in velocity controller and estimate the maximum exponential decay rate, which is essential for designing a MF-CBFs-based safe velocity controller, along with an uncertainty bound on the violation rate of the stability condition. Furthermore, based on the obtained decay rate and the associated uncertainty bound, we establish probabilistic safety guarantees within the safety constraint of MF-CBFs. Lastly, we validate the proposed framework on the UR5e collaborative robot by evaluating its safety performance in an obstacle avoidance scenario. We further analyze the conservatism of each safe controller under different decay rates and uncertainty bounds.


%% file: related_works.tex
\section{Related Works}\label{sec:related_works}
\subsection{Control Barrier Functions}
Replying on the set invariance approach, Control Barrier Functions (CBFs) \cite{Ames2019CBFtheoryandapplications} have become the promising and scalable schemes to ensure safety. Although CBFs have demonstrated significant success in certain robotic systems, particularly when combined with quadratic programming \cite{WOLF202218}\cite{Hadi2024}\cite{YKECC2024}, their effectiveness is highly sensitive to the model uncertainty. Inaccurate system knowledge can compromise the safety guarantees provided by CBFs. To resolve this issue, adaptive and worst-case robust approaches are integrated with CBFs \cite{lopez2020robust}\cite{JANKOVIC2018359}, which show the significant improvement of the safety guarantee. Recently, data-driven learning methods  \cite{ykgpracbf}\cite{Castaneda2021GPCBF} have been introduced to the CBFs-based control design to reduce unknown model uncertainty. Furthermore, the accurate knowledge of the system model is no longer needed in MF-CBFs \cite{molnar2021model}, relaxing the restriction on the application of CBFs.

\subsection{Learning Lyapunov Functions}
The significant demand for Lyapunov candidate synthesis arises from its crucial role in system stability analysis and within the framework of Lyapunov-based control \cite{guo_adaptive_2003}. While Lyapunov candidate identification has successfully proven in certain model-free contexts, parameter determination within these candidates remains challenging \cite{molnar2021model}, which motivates the recent research into data-driven Lyapunov candidate learning \cite{dai_lyapunov-stable_2021}. However, the finite number of snapshots prevents the learned Lyapunov candidate from ensuring certification guarantees of unseen samples, even those from an independent and identically distributed sample \cite{kenanian_data_2019}. The recent studies aim to alleviate this concern by seeking the bound of the confidence in generalizing the learned Lyapunov candidate \cite{boffi_learning_2021}.

The paper progresses as follows: Section~\ref{sec:preliminary_knowledges} introduces model-free CBFs as a preliminary before the primary framework is detailed in Section~\ref{sec:proposed_framework}. Then, Section~\ref{sec:experiment} presents the numerical example and the real-world experiment to validate the proposed framework. The conclusions and discussions are addressed in Section~\ref{sec:conclusions}. 

%% file: preliminary_knowledges.tex
\section{Preliminaries}\label{sec:preliminary_knowledges}
In this section, we first provide a brief review of Model-Free Control Barrier Functions (MF-CBFs), focusing on the  prerequisites for MF-CBFs-based control, particularly the Lyapunov function and its decay rate. Then, we establish the connection between MF-CBFs and the proposed method.

\subsection{Safety Guarantees with Model-Free CBFs}
Consider the following robot dynamics: 
\begin{equation}
M(\bm{q}) \ddot{\bm{q}} + C(\bm{q}, \dot{\bm{q}}) \dot{\bm{q}} + g(\bm{q}) = B\bm{u}, \label{pre:robot_dynamics}
\end{equation}
where $\bm{q} \in \mathbb{R}^n$ are robot joint positions in the configuration space, $\mathrm{Q} \subseteq \mathbb{R}^n$, and $ M(\bm{q}) \in \mathbb{R}^{n \times n}$ is the generalized mass matrix,  $ C(\bm{q}, \dot{\bm{q}}) \in \mathbb{R}^{n \times n}$ denotes Coriolis and centrifugal terms, $g(\bm{q}) \in \mathbb{R}^n$ includes the gravitational terms, and $B\in \mathbb{R}^{n\times m}$ is the system input matrix. Given that $h_q: \mathrm{Q} \rightarrow \mathbb{R}$ is a continuously differentiable, we define a safe set with $h_q$ as follows:
\begin{equation}
    \mathcal{S}_q = \{\bm{q} \in \mathrm{Q} \text{ }\vert\text{ } h_q(\bm{q}) \geq 0\}\label{pre:safe_set_confi}.
\end{equation}

We assume that $\big|\big|\frac{\partial h_q(\bm{q})}{\partial \bm{q}}\big|\big| \leq C_h$, $C_h>0$ for $\forall \bm{q} \in \mathcal{S}_q$, and there exists an exponentially stable velocity controller, $\bm{u} = k_q(\bm{q},\dot{\bm{q}})$, where $k_q:\mathrm{Q}\times \mathbb{R}^n \rightarrow \mathcal{U}$ and $\mathcal{U} \in \mathbb{R}^m$ is the input space, such that there exists a continuously differentiable Lyapunov function, $V(\bm{q},\dot{\bm{e}})$, satisfying the following conditions:
\begin{align}
   &k_1||\dot{\bm{e}}|| \leq V(\bm{q},\dot{\bm{e}}) \leq k_2||\dot{\bm{e}}||, k_1,k_2\in \mathbb{R}_{>0} \nonumber \\
   &\dot{V}(\bm{q}, \dot{\bm{e}},\dot{\bm{q}}, \ddot{\bm{q}}_d, \bm{u}) \leq -\lambda V(\bm{q},\dot{\bm{e}}), \label{pre:stable_vel_controller}
\end{align}
where $\dot{\bm{e}} = \dot{\bm{q}} - \dot{\bm{q}}_d$ denotes the velocity tracking error between $\dot{\bm{q}} $ and desired velocity, $\dot{\bm{q}}_d$. With the stable velocity controller, we design a safe controller based on the model-free control barrier functions in the following:

\begin{theorem}[\cite{molnar2021model}]
    Consider the robot system \eqref{pre:robot_dynamics} and the initial condition $(\bm{q}_0,\dot{\bm{e}}_0) \in \mathcal{S}_V$ defined by
    \begin{equation}
        \mathcal{S}_V = \{(\bm{q},\dot{\bm{e}}) \in \mathrm{Q} \times \mathbb{R}^n \text{ }:\text{ } h_V(\bm{q},\dot{\bm{e}}) \geq 0\} \label{pre:init_condition_set},
    \end{equation}
    where $h_V(\bm{q},\dot{\bm{e}}) = -V(\bm{q},\dot{\bm{e}}) + \alpha_e h_q(\bm{q})$ with $\alpha_e = \frac{k_1(\lambda -\alpha)}{C_h}$. For $\lambda > \alpha$, if there exists a safe velocity, $\dot{\bm{q}}_s$ satisfying
    \begin{equation}
        \frac{\partial h_q(\bm{q})}{\partial \bm{q}}\dot{\bm{q}}_s \geq -\alpha h_q(\bm{q}), \quad \forall\bm{q}\in \mathcal{S}_q, \label{pre:model_free_safety_condition}
    \end{equation}
    and a stable velocity controller tracking $\dot{\bm{q}}_s$, $\mathcal{S}_q$ is forward invariant and the system \eqref{pre:robot_dynamics} is \textit{safe} with respect to $\mathcal{S}_q$.
\end{theorem}

Note that the potential unknown disturbances in real robotic systems could deteriorate stability and safety guarantees. The effect of these disturbances can be addressed with the leverage of the input-to-state exponential stability (ISS) \cite{sontagISS2008} and safety (ISSf) \cite{ShishirISSfCBFs2019} conditions, which can be efficiently integrated with model-free CBFs. Let $\mathcal{S}_d$ be a larger safe set, e.g., $\mathcal{S}_d\supseteq \mathcal{S}_q$ . Define it as:
\begin{equation}
    \mathcal{S}_d = \{\bm{q} \in \mathrm{Q} \text{ }\vert\text{ }h_d(\bm{q})\geq 0\}, \label{pre:inflated_safe_set}
\end{equation}
where $ h_d(\bm{q}) = h_q(\bm{q}) + \gamma(||\bm{d}||_{\infty})$, and $\gamma(\cdot)$ is some class $\mathcal{K}$ functions. Then the following corollary ensures safety guarantees in the presence of disturbances.

\begin{corollary}[\cite{molnar2021model}]
    Consider the system with unknown bounded disturbance, $\bm{d} \in \mathbb{R}^m$,
    \begin{equation}
        M(\bm{q}) \ddot{\bm{q}} + C(\bm{q}, \dot{\bm{q}}) \dot{\bm{q}} + g(\bm{q}) = B(\bm{u}+\bm{d}), \label{pre:robot_dynamics_dis}
    \end{equation}
    and the initial condition, $(\bm{q}_0,\dot{\bm{e}}_0) \in \mathcal{S}_{V_d}$ defined by the inflated safe set:
    \begin{equation}
        \mathcal{S}_{V_d} = \{(\bm{q},\dot{\bm{e}}) \in \mathrm{Q} \times \mathbb{R}^n \text{ }:\text{ } h_{V_d}(\bm{q},\dot{\bm{e}}) \geq 0\}, \label{pre:inflated_init_condition_set}
    \end{equation}
    where $h_{V_d}(\bm{q},\dot{\bm{e}}) = h_V(\bm{q},\dot{\bm{e}}) +\gamma(||\bm{d}||_{\infty})$. Suppose that there exists a stable velocity controller satisfying the following ISS condition, 
    \begin{equation}
        \dot{V}(\bm{q}, \dot{\bm{e}},\dot{\bm{q}}, \ddot{\bm{q}}_d, \bm{u},\bm{d}) \leq -\lambda V(\bm{q},\dot{\bm{e}}) +\iota(||\bm{d}||_{\infty}), \label{pre:ISS_vel_ctrl}
    \end{equation}
    where $\iota$ is an extended class $\mathcal{K}^{e}_{\infty}$ function.
    For $\lambda > \alpha$, if the velocity controller \eqref{pre:ISS_vel_ctrl} tracks $\dot{\bm{q}}_s$ ensuring $ \frac{\partial h_q(\bm{q})}{\partial \bm{q}}\dot{\bm{q}}_s \geq -\alpha h_q(\bm{q}), \forall\bm{q}\in \mathcal{S}_d$,  the system \eqref{pre:robot_dynamics_dis} is \textit{ISSf} with respect to $\mathcal{S}_d$.
    \label{pre:ISSf-model_free_cbf}
\end{corollary}
\begin{remark}
    It is worth investigating the Lyapunov candidate function and its exponential decay rate in the robotic system before designing a safe velocity controller. For example, $\alpha$ from \eqref{pre:model_free_safety_condition} should be adjusted, satisfying $\lambda > \alpha$ at the same time. In the following, we introduce the framework for learning the Lyapunov candidate function from data (e.g. joint trajectories) with the uncertainty bound and subsequently estimating the exponential decay rate. Consequently, $\alpha$ can be determined in a constructive way.
\end{remark}

%% file: neural_network.tex
\begin{figure*}[t]  
    \centering
    \includegraphics[width=1\linewidth]{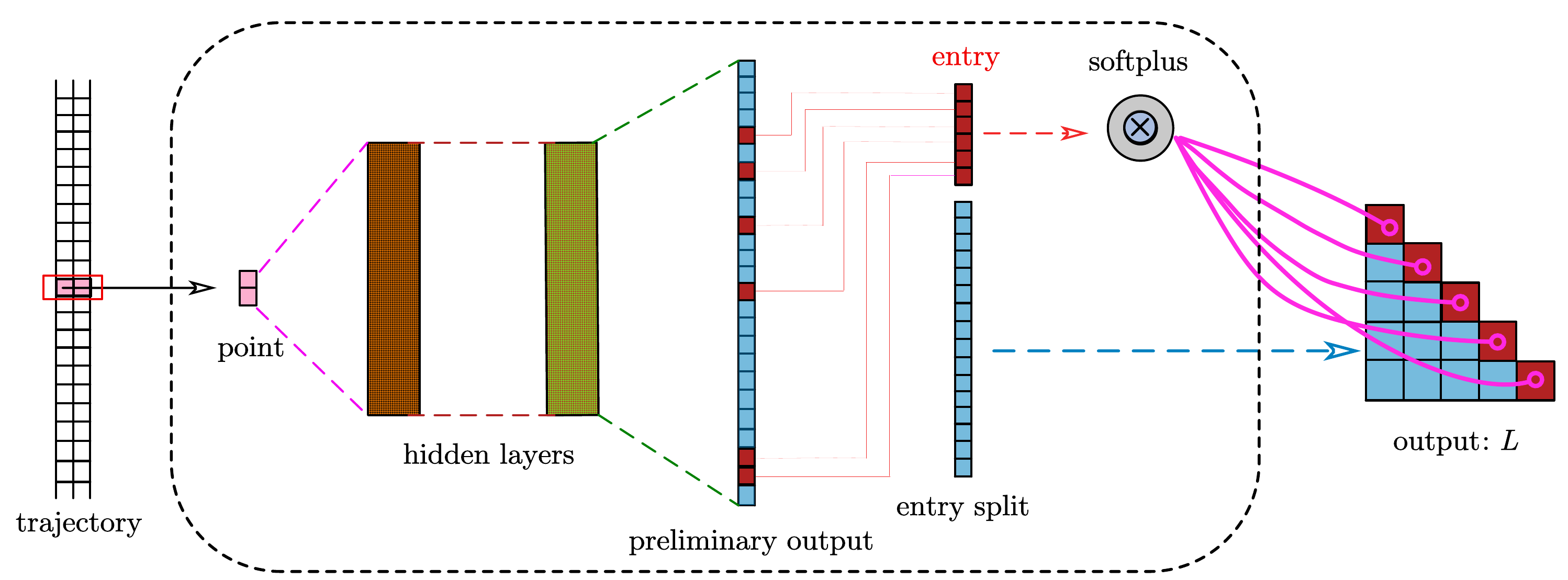}  
     \caption{\small{Neural Network in learning the Cholesky factor. The input is the current sampled tracking error trajectory and its derivative. The preliminary output consists of $\frac{n (n+1)}{2}$ scalars, where $ n $ of them are then gathered and input into \texttt{Softplus} to receive the positive diagonal entries. While the rest scalars in the preliminary output are directly adopted in the final output.}}  
     \vspace{-0.3cm}
    \label{fig:neural_network}
\end{figure*}

%% file: proposed_framework.tex
\section{Proposed Framework}\label{sec:proposed_framework}
This section presents a data-driven method for estimating an exponential decay rate of a dynamical system (e.g., a robot manipulator) by learning a Lyapunov function based on a Neural Network. Furthermore, we formally derive a generalization error bound for the learned Lyapunov function. Lastly, we show how the decay rate and the constraint violation parameter can be integrated into MF-CBFs through this probabilistic framework to ensure probabilistic safety guarantee.




%% file: method.tex
\subsection{Learning Lyapunov Function}
Lyapunov functions can be obtained using various optimization methods for systems with known dynamics; however, obtaining a Lyapunov function when the system dynamics are unknown remains a significant challenge. To ensure safety based on Corollary~\ref{pre:ISSf-model_free_cbf}, it is necessary to know the exponential decay rate $\lambda$.



To address this challenge, a data-driven approach is proposed to determine the exponential decay rate and to learn the Lyapunov candidate based on \cite{boffi_learning_2021}.

\subsection{Lyapunov Candidate Structure}
The candidate Lyapunov function, $V: \mathbb{R}^n \rightarrow \mathbb{R}$, is chosen to be a quadratic form:
\begin{equation}
    V(\bm{x}) = \bm{x}^{\top} P \bm{x},\label{eqn:Lyap_P}
\end{equation}
where $\bm{x} \in\mathbb{R}^{n}$ is the state, and $P\in\mathbb{R}^{n\times n}$ is a symmetric positive definite matrix $P=P^{\top}$. In the following,  a  Cholesky factorization given by $P=L L^{\top}$ is considered \cite{Hafstein2024}. The Cholesky factor $L$ of a positive definite matrix is a real lower triangular matrix with positive diagonal entries. Consequently, the positivity of the diagonal elements of $L$ will be enforced in the learning algorithm.

To show exponential stability and thus identify an exponential decay rate $\lambda$, the following should hold
\begin{align}
\dot{V}(\bm{x})\leq-\lambda V(\bm{x}).\label{eqn:exponential_stability}
\end{align}

Since the system dynamics is unknown, it is not possible to compute $\dot{V}(\bm{x})$ directly; thus, it will be approximated from data using the Neural Network presented in the following.


\subsection{Neural Network for Representation of Lyapunov Function}
The Cholesky factor $L$ is learned by a fully connected Neural Network with two hidden layers \cite{boffi_learning_2021}, each containing $32$ neurons, using  \texttt{ReLU} activation functions. The input to the Neural Network is a set of $N$ sampled state trajectories. The preliminary output consists of $\frac{n (n+1)}{2}$ scalars, which are the elements of the lower triangular matrix $L$. 

Different from the fully connected Neural Network applied in the previous work \cite{boffi_learning_2021}, the \texttt{Softplus} function, e.g., $\log(1 + e^t)$, is applied to the $n$ scalars among the preliminary output $\frac{n(n+1)}{2}$ scalars to ensure positive diagonal entries in the Cholesky factor $L$. The Neural Network structure is visualized in \figref{fig:neural_network} where the preliminary output is processed using \texttt{Softplus} and is reshaped to a lower triangular matrix, which is then multiplied by its transpose to obtain the unique corresponding positive definite matrix $P$.

The Neural Network is trained based on a loss defined from the constraint \eqref{eqn:exponential_stability}, i.e., the following function 
\begin{align}
    h(\bm{x},\dot{\bm{x}})=\dot{V}(\bm{x})+\lambda V(\bm{x})+\gamma,\label{eqn:constraint_h}
\end{align}
where $\gamma>0$. The parameter $\gamma$ is introduced to ensure that the constraint \eqref{eqn:exponential_stability} is strictly satisfied. This is necessary to determine an upper bound on the generalization error from the training data, see \cite{boffi_learning_2021}. 
Finally, the loss function is defined as
\begin{align}
    \psi(\bm{x}, \dot{\bm{x}})=\max\left\{0, h(\bm{x}, \dot{\bm{x}})\right\}.
\end{align}

The loss function is used for the learning, which is accomplished by back propagation. 
Specifically, the weights are carried over from the end of the current epoch to the next epoch, while their derivatives are reset at the beginning of each epoch \cite{Ruder2017}.


Pseudocode is provided in Algorithm~\ref{algorithm_1} to clarify how the Lyapunov function $V(\bm{x})$, the decay rate $\lambda$, and the constraint violation with respect to \eqref{eqn:exponential_stability}, $\epsilon$ are computed. The constraint violation $\epsilon$ is introduced to increase the practical applicability of the method. The value of $\epsilon$ is the allowed violation of the exponential decay condition, which is expected to be necessary for real-world systems. The value of $\epsilon$ is included in the safety condition to guarantee  safety.

\input{lambda_algorithm}


\subsubsection{Relation Between Amount of Data and Generalization Error}
The generalization error of the learned Lyapunov function, i.e. the probability of violating \eqref{eqn:exponential_stability} at new initial conditions, can be bounded from above based on e.g. Lemma~4.1 in \cite{boffi_learning_2021}. Such bounds depend on the number of trajectories used for training, the complexity of the function class and the value of $\gamma$. The training should rely on many trajectories, the class of functions should be simple, and $\gamma$ should be large. Since the bound is not constructive, we compute a bound of the generalization error based on test trajectories as described in Lemma~\ref{pro:classification_prob}.

\subsubsection{Numerical example} 
We present the process of learning a candidate Lyapunov function  from data on an illustrative example.
Consider a linear dynamical system,
\begin{equation}
    \begin{aligned}
        \dot{\bm{x}} &= \begin{bmatrix} 0 & 1 \\ -1 & -2 \end{bmatrix} \bm{x},\label{eqn:lin_sys}
\end{aligned}
\end{equation}
where $ \bm{x}\in\mathds{R}^2 $ is state. 
Algorithm~\ref{algorithm_1} is used to compute a Lyapunov function from $n_T=10$ state trajectories with different initial conditions. Both the state $\bm{x}$ and its derivative $\dot{\bm{x}}$ are used to train the Neural Network \cite{Ruder2017}, since they are used for computing the loss. We compute $\dot{\bm{x}}$ by numerical differentiation of $\bm{x}$. The Lyapunov candidate is given by:
\begin{equation}
    V(\bm{x}) = \bm{x}^{\top}  L  L^{\top} \bm{x},\label{eqn:Lyap_L}
\end{equation}
where $ L $ is a Cholesky factor. 
Algorithm~\ref{algorithm_1} determines the maximal exponential decay rate $\lambda$ via bisection. The maximal exponential decay rate for \eqref{eqn:lin_sys} is computed to $\lambda=1.5$ with the corresponding Cholesky factor
\begin{align}
L = \begin{bmatrix}
0.715 & 0 \\
0.609 & 0.0819
\end{bmatrix}.\label{eqn:L_val_sim}
\end{align}

\begin{figure}[t]
    \centering
    \includegraphics[width=1\linewidth]{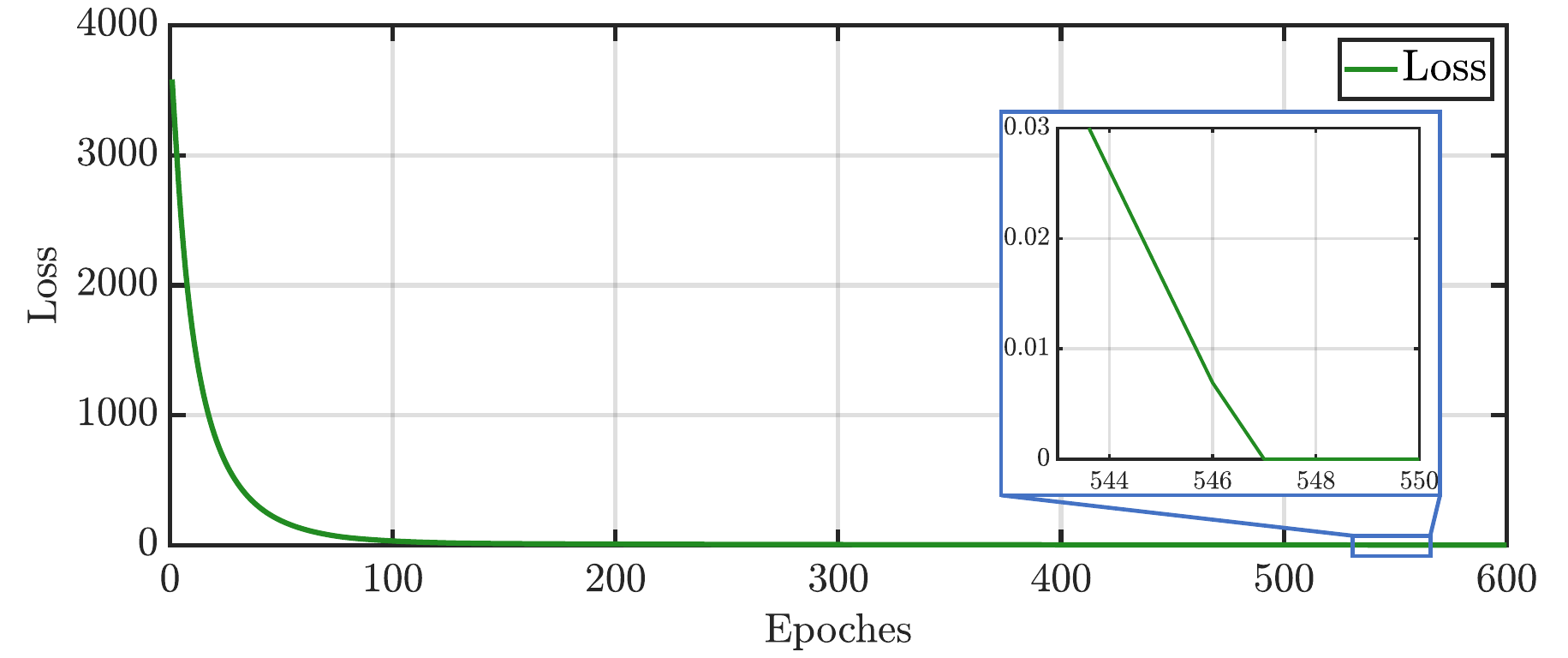}
    \caption{\small{ The training loss during each Epoch. The loss decreases to zero after 547 Epoches.}}
    \vspace{-0.3cm}
    \label{fig:sim_result}
\end{figure}

\figref{fig:sim_result} shows the training loss during the Neural Network training process. It is seen that the loss decreases to zero after 547 Epoches. For this example the underlying dynamical system is known; thus, we can check if the Cholesky factor in \eqref{eqn:L_val_sim} generates a valid Lyapunov function from \eqref{eqn:Lyap_L}. First, it is observed that $V(\bm{x})$ is positive definite as the defining matrix $P=LL^\top$ is positive definite. Second, it is seen that $\dot{V}(\bm{x})+\lambda V(\bm{x})$ is negative definite; thus, $\lambda$ is a valid decay rate of the considered system \eqref{eqn:lin_sys}.


\subsection{Probabilistic Safety Guarantees}
To use the learned value of the exponential decay rate in the MF-CBF, it is necessary to determine an upper bound of the generalization error. This bound is combined with MF-CBF to establish a probabilistic safety condition. 

To this end, classification approximation probability \cite{JMLR:v6:langford05a} is used to bound the violation rate of the following condition
\begin{align}
\dot{V}(\bm{x},\dot{\bm{x}}) + \lambda V(\bm{x}) - \epsilon \leq 0.
\label{pro:const_test1}
\end{align}

Specifically, we rely on the Chernoff bound \cite[Lemma~3.6]{JMLR:v6:langford05a} to bound the true rate of violating  \eqref{pro:const_test1}. The true rate of violation is denoted $c_D$ and is associated to a specific distribution of initial conditions $D$. Such bound is computed based on the number of constraint violations $\hat{c}_S$ from $m$ test samples (initial conditions) from distribution $D$. In the following, we consider constraint \eqref{pro:const_test1} to be a classifier with classes true and false.

\begin{lemma}[\cite{JMLR:v6:langford05a}] For all classifiers $c$, for all $\delta\in(0,1]$
\begin{align}
\underset{S\sim D^m}{Pr}
\left(c_D\leq\bar{c}_D\right)\geq1-\delta.
\end{align}
where
\begin{align}
\bar{c}_D=\frac{\hat{c}_S}{m}+\sqrt{\frac{\log(1/\delta)}{2m}}.
\end{align}
\label{pro:classification_prob}
\end{lemma}

The value of $\bar{c}_D$ provides an upper bound on the generalization error, i.e. an upper bound on the rate of violation of \eqref{pro:const_test1} along an entire trajectory from the test samples. Recall that the considered  MF-CBF conditions applies point-wise along a trajectory; hence, $\bar{c}_D$ will be used to generate a conservative bound on such point-wise condition.

Consider the safe set, \eqref{pre:inflated_safe_set} and the system, \eqref{pre:robot_dynamics_dis} and an exponentially stable velocity controller that satisfies the learned Lyapunov function and its estimated decay rate, $\lambda$. For any $\alpha$ such that $\alpha < \lambda$, the system is \textit{ISSf} with probability at least $1-\delta$ if there exist a safe velocity, $\dot{\bm{q}}_s$ satisfying \eqref{pro:epsilon_cond} below. 

Inspired by \cite{molnar2021model}, this probabilistic safety is obtained in the following. For the sake of brevity, $\dot{V}(\bm{q}, \dot{\bm{e}},\dot{\bm{q}}, \ddot{\bm{q}}_s, \bm{u},\bm{d}), {V}(\bm{q},\dot{\bm{e}}), \frac{\partial h_q(\bm{q})}{\partial \bm{q}}$ are denoted by $\dot{V}, V, \Delta h_q(\bm{q})$, respectively. Let us define ${h}_{V_d}(\bm{q},\dot{\bm{e}}) = -{V}+\alpha_e h_q(\bm{q}) + \gamma(||\bm{d}||_{\infty})$ from \eqref{pre:init_condition_set}, and its derivative is in the following:
\begin{align}
\dot{h}_{V_d}(\bm{q},\dot{\bm{e}}) &= -\dot{V}+ \alpha_e \Delta h_q(\bm{q})\dot{\bm{q}}
\end{align}
By Lemma~\ref{pro:classification_prob} with probability at least $1-\delta$
\begin{align}
-\dot{V}(\bm{x},\dot{\bm{x}}) \geq \lambda V(\bm{x}) - \epsilon
\end{align}
is violated with a rate lower than $\bar{c}_D$ for trajectories initialized in $D$.

Consequently, with probability at least $1-\delta$ the following inequality is violated with a rate lower than $\bar{c}_D$:
\begin{align}
\dot{h}_{V_d}(\bm{q},\dot{\bm{e}})&\geq \lambda V -\epsilon + \alpha_e \Delta h_q(\bm{q})\dot{\bm{q}} \nonumber \\
&\geq -\alpha h_{V_d}(\bm{q},\dot{\bm{e}}).
\label{pro:prob_Safety_cond}
\end{align}

We omit the detail proof between the first and second inequalities in \eqref{pro:prob_Safety_cond} since the proof is provided in \cite[Theorem3]{molnar2021model}. Consequently, we use $\epsilon$ in the safe velocity constraint, \eqref{pre:model_free_safety_condition} as follows:
\begin{equation}
        \frac{\partial h_q(\bm{q})}{\partial \bm{q}}\dot{\bm{q}}_s \geq -\alpha\big(h_q(\bm{q})- \epsilon_{c}\big) \quad \text{with } \epsilon_c = \frac{\epsilon}{\alpha}.
        \label{pro:epsilon_cond}
\end{equation}


%% file: lambda_algorithm.tex
\begin{algorithm}
    \caption{Binary Search for maximum $\lambda$, $\epsilon$, $V$}
    \begin{algorithmic}[1]
        \State \textbf{input:} $\lambda_{\min}$, $\lambda_{\max}$, $R,$ $n_T$ sampled trajectories
        \State $\lambda_{\text{best}} \gets \lambda_{\min}$
        \Repeat
            \State $\lambda \gets \frac{\lambda_{\min} + \lambda_{\max}}{2}$
            \State Train Neural Network with decay rate $\lambda$ with $n_T$ sampled trajectories
            \If {no train converges} 
                \State Reduce upper bound: $\lambda_{\max} \gets \lambda$
            \Else
                \State Update the corresponding $V$
                \State Update the best feasible $\lambda$: $\lambda_{\text{best}} \gets \lambda$
                \State Increase lower bound: $\lambda_{\min} \gets \lambda$
            \EndIf
        \Until {$\lambda_{\max} - \lambda_{\min} < R$} (resolution limit)
        \State Find the smallest non-negative $\epsilon$ such that all $n_T$ sampled trajectories satisfy:
        \State $ \dot{V} + \lambda V - \epsilon \leq 0 $
        \State \Return $\lambda_{\text{best}}, \epsilon, V$
    \end{algorithmic}
    \label{algorithm_1}
\end{algorithm}

%% file: experiment.tex
\section{Experimental validation}\label{sec:experiment}
\begin{figure}[!htb]
\centering
  \def\svgscale{1}
\graphicspath{{figures/}}
{\footnotesize
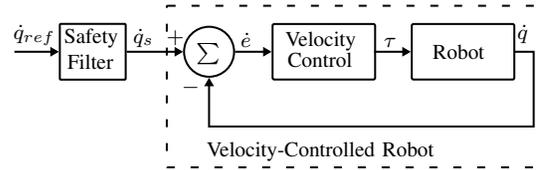}
\caption{\small{Block diagram of a safe velocity control via the velocity-controller robot, UR5e.}}\label{fig:blockRobotPID}
\end{figure}

This section presents the experimental results to validate the proposed framework. A learned Lyapunov function is obtained using a Neural Network trained on data from a UR5e manipulator, as shown in \figref{fig:blockRobotPID}, which is controlled by its built-in velocity controller. Using the decay rate $\lambda$ and the parameter $\epsilon$ associated with the learned Lyapunov function, a safety filter based on MF-CBFs is deployed. Several cases with varying design parameters for the obstacle avoidance scenario are subsequently examined. The robot is controlled and interfaced via the UR\_RTDE library \cite{Anders_UR_RTDE}.

\subsection{Learning Lyapunov Function}
To learn a Lyapunov function, the state is defined as $\bm{x}=(\bm{q},\dot{\bm{e}})$, and for illustrative purposes only two joints are considered. To train the Neural Network representing the Lyapunov function, reference angular velocities are given to the considered joints which start from the initial velocities \( \dot{\bm{q}}_0 \sim \text{Unif}([0,1] \times [0,1]) \).
The resulting angular velocity $ \dot{\bm{q}} $, along with the reference velocity $ \dot{\bm{q}}_{ref} $, are recorded and used to calculate the velocity tracking error,
\begin{equation}
    \bm{\dot{e}} = \dot{\bm{q}}_{ref} - \dot{\bm{q}},
\end{equation}
from which the 
Lyapunov candidate below is constructed \cite{molnar2021model}:
\begin{equation}
    V = \bm{\dot{e}}^{\top} P \bm{\dot{e}}.
\end{equation}
\begin{figure}[t]
    \centering
    \includegraphics[width=1\linewidth]{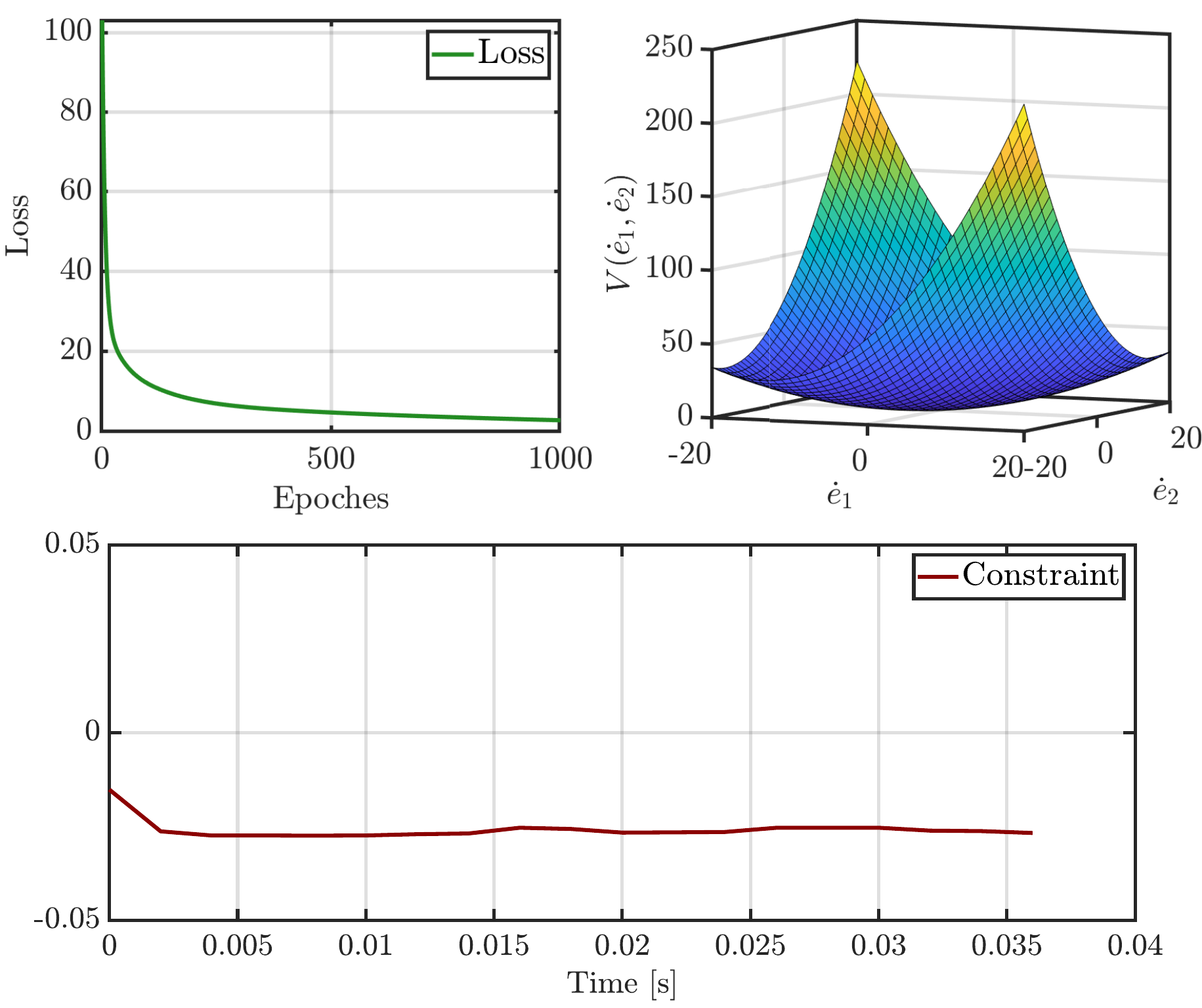}  
    \caption{\small{The result from the data of UR5e. The training loss during each Epoch is on the left. The resulting Lyapunov function $V(\bm{x}$ is on the right. The time-specified constraint, $\dot{V}(\bm{x})+\lambda V(\bm{x})$ from an initial condition is displayed in the bottom. }}  
    \vspace{-0.3cm}
    \label{fig:real_result}
\end{figure}
Note that in the experiment, noise is observed in the state, which can inevitably undermine the learned stability proof. With this concern, the stability proof is relaxed by $\epsilon$, found in Algorithm~\ref{algorithm_1}.

The output from the Neural Network with 99 training samples is
\begin{equation}
    P = LL^{\top}, \quad L =
    \begin{bmatrix}
        0.418 & 0 \\
        -0.281 & 0.259
    \end{bmatrix},
\end{equation}
which corresponds to minimum \( \epsilon\) and the behavior of the Lyapunov candidate visualized in \figref{fig:real_result}, where, in the bottom, the constraint history is replayed on a training set already learned by the Neural Network. In general, the Lyapunov candidate retains the exponential decay property even in the presence of noise, and the constraint on exponential decay remains satisfied.

The true rate of violating the constraint (generalization error) is bounded by \(7.39\%\) from Lemma~\ref{pro:classification_prob}, given that \(\delta=0.01\) and driven by 499 unseen training samples. Building on the obtained minimum $\epsilon = 0.0274$ and maximum $\lambda = 5$, the next section discusses the associated safety guarantee with probabilistic confidence.

\begin{figure}[t]
    \centering
\includegraphics[width=1\linewidth]{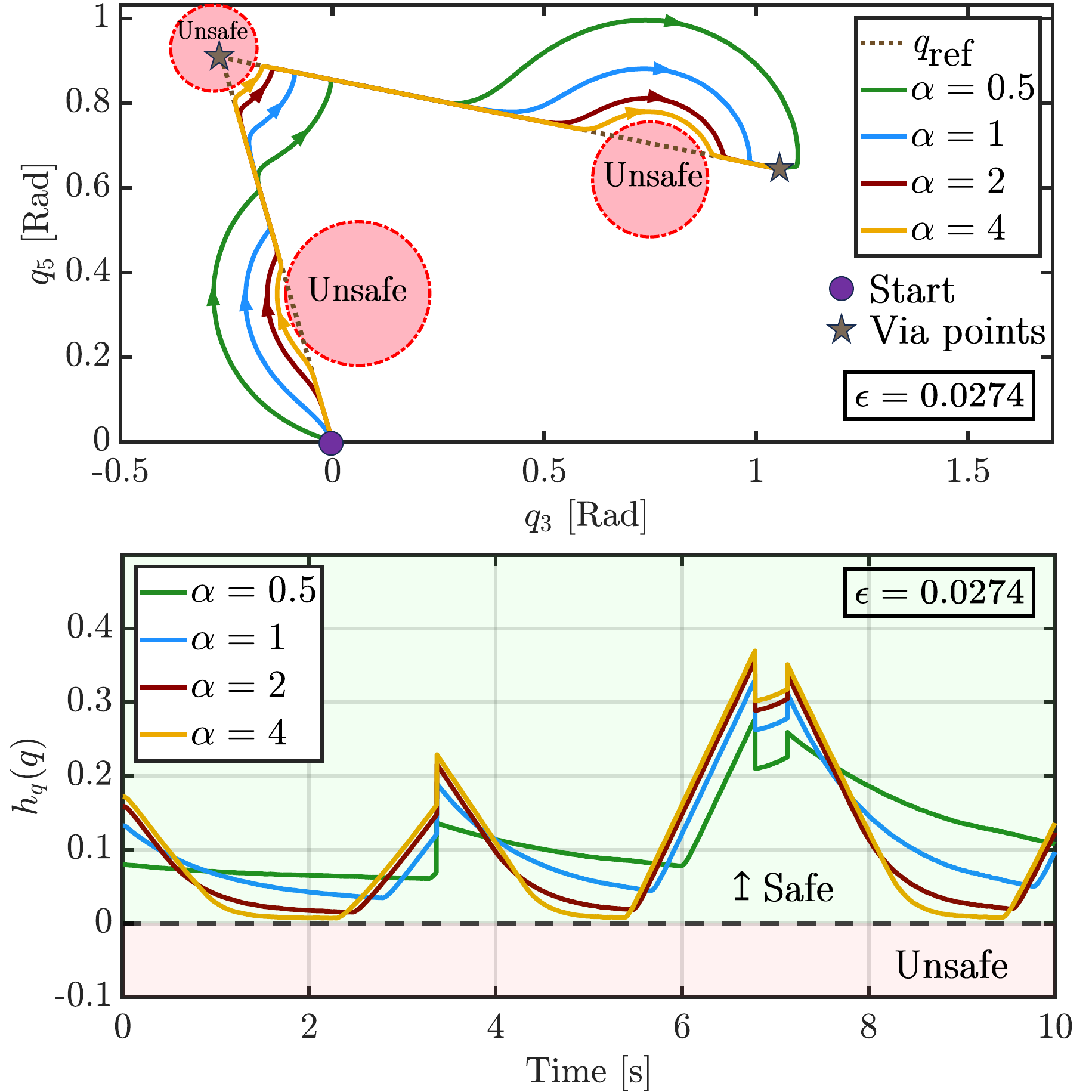}
    \caption{\small{The performance of a safe velocity controller with $\epsilon =  0.0274$ and different $\alpha$ such that $0 < \alpha < \lambda = 5$ provided by Algorithm~\ref{algorithm_1}. The first row represent the scenario where a robot avoid the unsafe area, and the second one shows safety guarantees, $h_q({q}) \geq 0$ in all cases.}}
    \vspace{-0.3cm}
    \label{fig:results}
\end{figure}
\subsection{Demonstration of Safety Guarantees on Real Robot}
We conduct the real-robot experiments to validate the proposed method, demonstrating it in the joint space scenario,  where joints should avoid unsafe regions as shown on the top of \figref{fig:results}. We employ wrist joints (e.g. $q_3$ and $q_5$) of UR5e in this experiment and provide the desired joint positions, $\bm{q}_{ref}$ along with a linear trajectory. A proportional gain is applied to convert the desired joint positions into velocity commands as input for the robot, and we define each safety function corresponding to $q_3, q_5$, by $h_q(q) = l - r $ where $l$ represents the distance from each obstacle and $r$ indicates each radius, following \cite{molnar2021model}. Subsequently, a safe velocity controller is also designed in the similar to \cite[Example 1]{molnar2021model}, but with various $\alpha$ values, constrained by $0 < \alpha < \lambda  = 5$. Lastly, we enforce the probabilistic safety condition \eqref{pro:epsilon_cond} to ensure safety guarantees and validate the safe controller in the scenario.

Fig.~\ref{fig:results} shows the performance of the safe velocity controller. We intentionally choose the reference positions that invade unsafe regions in order to compare each performance with the different $\alpha$ values such as $\alpha = 0.5, 1, 2, 4$ and $\epsilon = 0.0274$. Overall, the robot does not go into unsafe areas in red circles, which indicates that safety guarantees are successfully ensured. This is aligned with the minimum safety function, $h_q({q})$ associated with $q_3,q_5$ is always non-negative in all cases. It is observed that the conservatism decreases as $\alpha$ increases, and vice versa. To be a more general analysis of conservatism, we also conduct several cases with different $\epsilon$ values. Table.~\ref{table_1} shows the minimum $h_q({q})$ values, which allows us to analyze the conservatism clearly, for example, the larger $h_q({q})$ is, the more conservative it becomes. It is shown that the conservativeness increases as $\epsilon$ increases, which is because $\epsilon$ enforces the robustness against the uncertainty of exponential stability in \eqref{pro:epsilon_cond}. For comparison, when we choose $\alpha$ based on the standard condition, $\alpha>0$ such as $10,20,30$ without any $\epsilon$, the safety violation occurs.

 The results represent that there exists a trade-off relationship between conservatism and the selections of $\alpha, \epsilon$. Furthermore, it is worth noting that there exists a compromise between  $\alpha$ and $\epsilon$ each other. For example, the $h_q(q)$ is $0.0178$ with $\epsilon = 0.07$ and $\alpha = 4$, however, the level of conservatism remains similar when we choose  $\epsilon = 0.274$ and $ \alpha = 2$. As a result, as long as $\alpha$ satisfies $0 < \alpha < \lambda $ and we know the minimum $\epsilon$ from Algorithm~\ref{algorithm_1}, any values of $\alpha$ can be freely chosen to satisfy the level of conservatism that applications require. 

\begin{table}[h] 
\centering
\input{table_1}
\caption{\small{Analysis of conservatism with minimum safety function $h_q(q)$. We compare between the selected $\alpha$ values from the proposed method \textbf{(Ours)} and randomly chosen ones under the standard condition $(\alpha>0, \epsilon = 0)$, along with their performances corresponding to various $\epsilon$ cases. }}
\label{table_1}
\vspace{-0.3cm}
\end{table}

%% file: figures/blockRobotPID.pdf_tex
\begingroup%
  \makeatletter%
  \providecommand\color[2][]{%
    \errmessage{(Inkscape) Color is used for the text in Inkscape, but the package 'color.sty' is not loaded}%
    \renewcommand\color[2][]{}%
  }%
  \providecommand\transparent[1]{%
    \errmessage{(Inkscape) Transparency is used (non-zero) for the text in Inkscape, but the package 'transparent.sty' is not loaded}%
    \renewcommand\transparent[1]{}%
  }%
  \providecommand\rotatebox[2]{#2}%
  \newcommand*\fsize{\dimexpr\f@size pt\relax}%
  \newcommand*\lineheight[1]{\fontsize{\fsize}{#1\fsize}\selectfont}%
  \ifx\svgwidth\undefined%
    \setlength{\unitlength}{200.12654114bp}%
    \ifx\svgscale\undefined%
      \relax%
    \else%
      \setlength{\unitlength}{\unitlength * \real{\svgscale}}%
    \fi%
  \else%
    \setlength{\unitlength}{\svgwidth}%
  \fi%
  \global\let\svgwidth\undefined%
  \global\let\svgscale\undefined%
  \makeatother%
  \begin{picture}(1,0.30822698)%
    \lineheight{1}%
    \setlength\tabcolsep{0pt}%
    \put(0.78788972,0.20550397){\color[rgb]{0,0,0}\makebox(0,0)[lt]{\lineheight{1.25}\smash{\begin{tabular}[t]{l}Robot\end{tabular}}}}%
    \put(0.34227176,0.20713014){\color[rgb]{0,0,0}\makebox(0,0)[lt]{\lineheight{1.25}\smash{\begin{tabular}[t]{l}$\sum$\end{tabular}}}}%
    \put(0,0){\includegraphics[width=\unitlength,page=1]{blockRobotPID.pdf}}%
    \put(0.69585341,0.22942501){\color[rgb]{0,0,0}\makebox(0,0)[lt]{\lineheight{1.25}\smash{\begin{tabular}[t]{l}$\tau$\end{tabular}}}}%
    \put(0,0){\includegraphics[width=\unitlength,page=2]{blockRobotPID.pdf}}%
    \put(0.22375784,0.23692015){\color[rgb]{0,0,0}\makebox(0,0)[lt]{\lineheight{1.25}\smash{\begin{tabular}[t]{l}$\dot{q}_s$\end{tabular}}}}%
    \put(0.42614737,0.22942501){\color[rgb]{0,0,0}\makebox(0,0)[lt]{\lineheight{1.25}\smash{\begin{tabular}[t]{l}$\dot{e}$\end{tabular}}}}%
    \put(0.94701262,0.24441558){\color[rgb]{0,0,0}\makebox(0,0)[lt]{\lineheight{1.25}\smash{\begin{tabular}[t]{l}$\dot{q}$\end{tabular}}}}%
    \put(0.3156419,0.14214479){\color[rgb]{0,0,0}\makebox(0,0)[lt]{\lineheight{1.25}\smash{\begin{tabular}[t]{l}$-$\end{tabular}}}}%
    \put(0.28691366,0.23629799){\color[rgb]{0,0,0}\makebox(0,0)[lt]{\lineheight{1.25}\smash{\begin{tabular}[t]{l}$+$\end{tabular}}}}%
    \put(0,0){\includegraphics[width=\unitlength,page=3]{blockRobotPID.pdf}}%
    \put(0.0940399,0.23527653){\color[rgb]{0,0,0}\makebox(0,0)[lt]{\lineheight{1.25}\smash{\begin{tabular}[t]{l}Safety \end{tabular}}}}%
    \put(0,0){\includegraphics[width=\unitlength,page=4]{blockRobotPID.pdf}}%
    \put(0.09992484,0.18725975){\color[rgb]{0,0,0}\makebox(0,0)[lt]{\lineheight{1.25}\smash{\begin{tabular}[t]{l}Filter\end{tabular}}}}%
    \put(0,0){\includegraphics[width=\unitlength,page=5]{blockRobotPID.pdf}}%
    \put(-0.00163703,0.24441544){\color[rgb]{0,0,0}\makebox(0,0)[lt]{\lineheight{1.25}\smash{\begin{tabular}[t]{l}$\dot{q}_{ref}$\end{tabular}}}}%
    \put(0,0){\includegraphics[width=\unitlength,page=6]{blockRobotPID.pdf}}%
    \put(0.36262393,0.02207853){\color[rgb]{0,0,0}\makebox(0,0)[lt]{\lineheight{1.25}\smash{\begin{tabular}[t]{l}Velocity-Controlled Robot\end{tabular}}}}%
    \put(0,0){\includegraphics[width=\unitlength,page=7]{blockRobotPID.pdf}}%
    \put(0.51063503,0.2302416){\color[rgb]{0,0,0}\makebox(0,0)[lt]{\lineheight{1.25}\smash{\begin{tabular}[t]{l}Velocity\end{tabular}}}}%
    \put(0.51459948,0.19120732){\color[rgb]{0,0,0}\makebox(0,0)[lt]{\lineheight{1.25}\smash{\begin{tabular}[t]{l}Control\end{tabular}}}}%
  \end{picture}%
\endgroup%

%% file: table_1.tex
\centering
\centering
\begin{tabular}{ccccccc}
\hline \hline
\multicolumn{2}{c|}{\multirow{2}{*}{Parameters}}                                                                                      & \multicolumn{5}{c}{$\epsilon$}                                                                                                                                                        \\ \cline{3-7} 
\multicolumn{2}{c|}{}                                                                                                                 & \multicolumn{1}{c|}{$0.0274$}          & \multicolumn{1}{c|}{$0.04$}            & \multicolumn{1}{c|}{$0.06$}            & \multicolumn{1}{c|}{$0.07$}            & $0.08$            \\ \hline
\multicolumn{1}{c|}{\multirow{4}{*}{\begin{tabular}[c]{@{}c@{}}$\alpha < \lambda$ \\ \textbf{(Ours)}\end{tabular}}} & \multicolumn{1}{c|}{$0.5$} & \multicolumn{1}{c|}{$0.0885$}          & \multicolumn{1}{c|}{$0.1059$}          & \multicolumn{1}{c|}{$0.1349$}          & \multicolumn{1}{c|}{$0.1502$}          & $0.1643$          \\
\multicolumn{1}{c|}{}                                                                                    & \multicolumn{1}{c|}{$1$}   & \multicolumn{1}{c|}{$0.0417$}          & \multicolumn{1}{c|}{$0.0522$}          & \multicolumn{1}{c|}{$0.0695$}          & \multicolumn{1}{c|}{$0.0785$}          & $0.0873$          \\
\multicolumn{1}{c|}{}                                                                                    & \multicolumn{1}{c|}{$2$}   & \multicolumn{1}{c|}{$0.0171$}          & \multicolumn{1}{c|}{$0.0229$}          & \multicolumn{1}{c|}{$0.0320$}          & \multicolumn{1}{c|}{$0.0368$}          & $0.0416$          \\
\multicolumn{1}{c|}{}                                                                                    & \multicolumn{1}{c|}{$4$}   & \multicolumn{1}{c|}{\bm{$0.007$}} & \multicolumn{1}{c|}{\bm{$0.0104$}} & \multicolumn{1}{c|}{\bm{$0.015$}} & \multicolumn{1}{c|}{\bm{$0.0178$}} & \bm{$0.0203$} \\ \hline 
                                                                                                         &                            &                                        &                                        &                                        &                                        &                   \\ \hline
\multicolumn{1}{c|}{\multirow{4}{*}{\begin{tabular}[c]{@{}c@{}}$\alpha > 0$, \\ $\textcolor{red}{\epsilon = 0}$ \end{tabular}}}                                                         & \multicolumn{1}{c|}{$10$}  & \multicolumn{2}{c}{\textcolor{red}{\bm{$-0.0011$}}}                                                            & -                                      & -                                      & -                 \\
\multicolumn{1}{c|}{}                                                                                    & \multicolumn{1}{c|}{$20$}  & \multicolumn{2}{c}{\textcolor{red}{\bm{$-0.0026$}}}                                                            & -                                      & -                                      & -                 \\
\multicolumn{1}{c|}{}                                                                                    & \multicolumn{1}{c|}{$30$}  & \multicolumn{2}{c}{\textcolor{red}{\bm{$-0.0028$}}}                                                            & -                                      & -                                      & -                 \\
\multicolumn{1}{c|}{}                                                                                    & \multicolumn{1}{c|}{$50$}  & \multicolumn{2}{c}{\textcolor{red}{\bm{$-0.0026$}}}                                                            & -                                      & -                                      & -                 \\ \hline\hline
\end{tabular}

%% file: conclusions.tex
\section{Conclusion}\label{sec:conclusions}
This paper proposes a framework for safe control using model-free CBFs with a data-driven approach. The proposed framework effectively learns the Lyapunov candidate, along with the decay rate, from the real experimental data. With the learned decay rate and the uncertainty bound on the violation rate of the stability condition, MF-CBFs have been accordingly designed with an incorporated probabilistic safety condition, successfully ensuring the safety in the robot tracking control. We expect that the proposed method enhances the usability of model-free CBFs to ensure safety guarantees, as it eliminates the need for manual tuning. Future research could focus on determining individual decay rates for each dimension to enhance flexibility and accuracy. Additionally, deriving a bound on the generalization error specific to this setting remains an open challenge.

\section*{Acknowledgement}
This work has been carried out within the framework of the EUROfusion Consortium, funded by the European Union via the Euratom Research and Training Programme (Grant Agreement No 101052200 — EUROfusion). Views and opinions expressed are however those of the authors only and do not necessarily reflect those of the European Union or the European Commission. Neither the European Union nor the European Commission can be held responsible for them. 
The work was also supported by Fabrikant Vilhelm Pedersen og Hustrus Legat.